\documentclass[useAMS,usenatbib]{mn2e}
\usepackage{epsf}
\usepackage{amssymb}
\usepackage{epsfig}
\usepackage[usenames]{color}
\usepackage{float}
\title[Observations UCXBs]{X-ray diagnostics of chemical composition of the accretion disk and donor star in UCXBs II: {\it XMM-Newton} observations }
\author[F.Koliopanos, M.Gilfanov, L.Bildsten, M. D{\'{\i}}az Trigo]{Filippos Koliopanos$^{1}$\thanks{filippos@mpa-garching.mpg.de}, Marat Gilfanov$^{1,2}$, Lars Bildsten$^3$, Maria D{\'{\i}}az Trigo$^4$  \\
$^{1}$MPI f\"ur Astrophysik, Karl-Schwarzschild str. 1, Garching, 85741, Germany\\
$^{2}$Space Research Institute of Russian Academy of Sciences, Profsoyuznaya 84/32, 117997 Moscow, Russia\\
$^3$Kavli Institute for Theoretical Physics, University of California, Santa Barbara, CA 93106-4030, USA\\
$^4$ESO, Karl-Schwarzschild-Strasse 2, 85748, Garching bei M\"unchen, Germany}
\begin{document}

\date{Accepted 2014 May 26. Received 2014 May 23; in original form 2014 April 2}

\pagerange{\pageref{firstpage}--\pageref{lastpage}} \pubyear{2014}

\maketitle

\label{firstpage}

\begin{abstract}
We search for the Fe ${\rm K\alpha}$ line in spectra of Ultra Compact X-ray Binaries (UCXBs). For this purpose
we have analyzed {\it XMM-Newton} observations of five confirmed UCXBs. 
We find that the object 2S 0918-549 -- whose optical 
spectrum bears tentative signatures of a C/O accretion disk -- is devoid of any emission 
features in the 6-7\,keV range, with an upper limit of less than 10\,eV
for the equivalent width (EW) of the iron line. 4U 1916-05 -- 
whose optical spectrum is consistent with reflection from a He-rich accretion disk -- 
exhibits a bright broad iron emission line. 
This behavior is in agreement with the theoretical predictions presented in \cite{2013MNRAS.432.1264K}.
Namely, we expect strong suppression of the Fe ${\rm K\alpha}$  emission line in spectra originating in moderately bright ($\rm LogL_x$ less than $\approx 37.5$)
UCXBs with C/O or O/Ne/Mg-rich donors. On the other hand the EW of the iron line in spectra from UCXBs with He-rich donors 
is expected to retain its nominal value of ${\rm \approx 100\,eV}$. Our analysis also reveals
a strong Fe ${\rm K\alpha}$ line in the spectrum of 4U 0614+091. This detection points towards a He-rich donor and seems to be at odds
with the source's classification as C/O-rich. Nevertheless, a He-rich donor would explain the bursting activity reported for this system. 
Lastly, based on our theoretical predictions, we attribute the lack of a strong iron emission line
-- in the two remaining UCXB sources in our sample (XTE J1807-294, 4U 0513-40) -- as an indication of a C/O or O/Ne/Mg white dwarf donor.
From the upper limits of the Fe ${\rm K\alpha}$ line EW in 4U 0513-40, 2S 0918-549 and XTE J1807-294 
we obtain a lower limit on the oxygen-to-iron ratio, O/Fe$\ge10\times{\rm[O/Fe]_{\odot}}$.

\end{abstract}


\section{Introduction}
Low mass X-ray binaries (LMXBs) with orbital periods of less than one hour are known as ultra-compact X-ray binaries. Their short orbital periods
imply orbits that are so tight that only an evolved compact donor could fit (e.g. \citealt{1984ApJ...283..232R}; \citealt*{1986ApJ...311..226N}).
Therefore, they must consist of a white dwarf  or a 
helium star that has filled its Roche lobe and is accreting onto
a neutron star \citep*[e.g.][]{1993ARep...37..411T, 1995ApJS..100..233I, 1995xrbi.nasa..457V, 2003ApJ...598.1217D, 2005ApJ...624..934D}. 

X-ray radiation from LMXBs usually consists of a primary and a reflected component \citep[e.g.][and references therein]{2010LNP...794...17G}. Primary 
radiation is most likely created in a hot optically thin corona, the disk itself or -- in the case of a neutron star accretor -- 
in the boundary layer that forms on the surface of the star.
The reflected component is produced when primary radiation is reprocessed  by the optically thick Shakura - Sunyaev accretion disk and by the surface of the 
donor star facing the compact object. X-ray reflection spectra originating in normal LMXBs with main sequence or red giant 
donors are characterized by a bright iron  $\rm K{\rm\alpha}$ emission 
line at $\approx 6.4-6.9$\,keV with an equivalent width (EW) typically of the order of $\approx100$\,eV  \citep[e.g.][]{2010ApJ...720..205C}. 

The composition of the accreting material in UCXBs is  expected to be significantly different from the solar 
composition accretion disks of typical LMXBs with main sequence or red giant donors.
Due to the nature of their compact donor, 
their chemical composition is expected to be consistent with the ashes of H burning (mostly He and $\rm ^{14}N$), 
He burning (mostly C/O) or carbon burning (mostly O/Ne). 
Depending on initial parameters and the environment (e.g. being part of a globular cluster) of UCXB progenitors they
will follow different evolutionary channels, resulting in a variety of
donors ranging from non-degenerate He stars to C-O or O-Ne-Mg white dwarfs \citep*[e.g.][]{1986A&A...155...51S,2002ApJ...565.1107P,2002A&A...388..546Y, 2004ApJ...607L.119B}.
Due to the fact that the different UCXB formation channels lead to degenerate donors of similar mass,  
determining the chemical composition of the disk (and therefore the donor star) in UCXBs can provide valuable insights 
into the evolutionary path that created each system.

In principle, a straight forward determination of the chemical composition of the disk and donor star in these systems could be achieved
using optical spectroscopy. A He-rich object could be identified by the presence of strong He lines in its spectrum 
\citep[e.g][]{2006MNRAS.370..255N}, while a C/O-rich object can be inferred by the lack of H and He lines combined with the presence 
of strong C and O lines \citep[e.g.][]{2004MNRAS.348L...7N, 2006A&A...450..725W}. However, due to their small sized accretion disks 
\citep{1994A&A...290..133V} the optical counterparts of UCXBs are quite faint, with V-band absolute magnitudes that are usually larger than 
$\approx$5 with distances ranging from $\approx$3-12\,kpc \citep[e.g.][]{2004MNRAS.348L...7N, 2006MNRAS.370..255N}. 
Therefore, ensuring definitive proof of the donor star composition -- using optical spectroscopy -- is a difficult task 
that can only be attempted using the latest generation of $>$8m telescopes.

In the case of X-ray spectroscopy the presence of O and Ne emission features --
that appear in the spectra of reprocessed emission from the accretion disk and 
white dwarf surface --  \cite[e.g.][]{2010MNRAS.407L..11M} and K-edges stemming from
absorbing material in the vicinity of the disk \cite[e.g.][]{2010ApJ...725.2417S}
could also provide direct indication of a C/O or O/Ne-rich disk and donor star.
However, due to increased interstellar absorption below 1\,keV and 
contamination of the reflected component by the primary emission, detection of these features with sufficient 
accuracy, often proves to be difficult. 
On the other hand, in \cite{2013MNRAS.432.1264K}, we demonstrated that 
the most striking and readily observable consequence of an anomalous C/O abundance 
involves the iron $\rm K{\rm\alpha}$ line located at 6.4\,keV. In particular, for a source of moderate 
luminosity (${\rm L}\rm_X\lesssim$ a few $10^{37}\rm erg\,s^{-1}$) we predicted a strong 
suppression of the Fe $\rm K{\rm\alpha}$ line in the case of a C/O or O/Ne/Mg WD donor. This translates to a more than an order of magnitude
decrease of the EW of the line.
On the other hand, in the case of a He-rich donor the iron line is expected to remain unaffected
with its EW similar to that observed in LMXBs with main sequence or red giant donors. As was demonstrated in Koliopanos et al.
these results are luminosity dependent. Namely,
for luminosities exceeding $\rm LogL\rm_X\approx 37.5 $, we expect C, O and Ne to be fully ionized in the inner parts of
the disk and thus canceling their screening effect on the iron line.

In addition to spectroscopic analysis, one could indirectly infer the accretion disk and donor star composition by 
studying a system's bursting activity. Gradual accumulation of H and/or He on the surface of an accreting neutron star can
eventually result in the ignition of the accumulated shell, producing a thermonuclear flash that is known as a type I X-ray burst
(e.g. \citealt{1976ApJ...205L.127G}; \citealt{1975ApJ...195..735H} and for a detailed review \citealt{2006csxs.book..113S}). Half of the 
total population of known UCXBs have exhibited bursting activity, ranging from a few sporadic bursts to frequent bursting activity with a recurrence
time extending from a few hours to a few weeks. Sporadic bursts could be due to trace amounts of H and He in an otherwise C/O-rich accreted material.
Frequent bursting activity, on the other hand, would require copious amounts of H and/or He to refuel the bursts. Consequently,
such an activity would support arguments in favor of a He-rich donor in a particular UCXB. 
This is illustrated by the detection of frequent burster \citep{2008ApJS..179..360G} 4U 1916-05 (discussed in this paper)
which is also an optically confirmed He-rich source
\citep{2006MNRAS.370..255N}. On other hand the same approach can yield conflicting results as is the case of
4U 0614+091 (also discussed in this work) whose bursting activity \citep{2010A&A...514A..65K, 2012ApJ...760..133L} seems to be inconsistent
with the strong evidence in favor of a C/O-rich donor \citep{2004MNRAS.348L...7N, 2006A&A...450..725W}.

In the present paper we investigate the chemical composition of the accretion disk in five UCXBs using X-ray spectroscopy.
In particular we analyze {\it XMM-Newton} spectra of these sources and compare our results with the findings of 
Koliopanos et al. in order to put a constraint on the 
chemical composition of  their accretion disks and donor star. We also analyze the spectra 
of two normal LMXBs, which we use as a control sample. In Section 2 
we present the sample of UCXBs and LMXBs chosen for our analysis. 
We describe the details of data extraction, report on the specifics of each 
observation and present our data analysis where we look 
for the existence and strength of a potential iron $\rm K{\rm\alpha}$ line at $\approx 6.4$\,keV. 
The analysis is followed by discussion and conclusions in Sections 3 and 4.

\section{OBSERVATIONS, DATA ANALYSIS AND results}

\begin{table*}
 \centering
 \begin{minipage}{115mm}
 \caption{List of {\it XMM-Newton} observations}
 \label{tab:obs}
 \begin{tabular}{@{}lcccc}
  \hline
   Object & obsID  & Date &Duration$^{1}$ (s)  &  Net count rate$^{2}$ (c/s) \\
     \hline
  UCXBs      &    &    &  \\
 \hline
 4U 0513-40       & 0151750101  &  2003-04-01  & 16420 & $35.40\pm0.05$\\ 
 4U 0614+091      & 0111040101  &  2001-03-13  & 13140 & $252.1\pm0.17$\\ 
 2S 0918-549      & 0061140101  &  2001-05-05  & 38070 & $60.47\pm0.06$\\ 
 XTE J1807-294    & 0157960101  &  2003-03-22  & 9293 & $35.42\pm0.11$\\ 
 4U 1916-05       & 0085290301  &  2002-09-25  & 14820 & $55.47\pm0.11$\\ 
      \hline
 LMXBs       &    &    & & \\
 \hline
 4U 1705-44       & 0402300201  &  2006-08-26  & 34130 & $27.23\pm0.05$\\
 SAX J1808.4-3658 & 0560180601  &  2008-10-01  & 45050 & $300.3\pm0.09$ $^{3}$\\    
  \hline
\end{tabular}

 \medskip
{$^{1}$Duration of filtered pn observation.\\
 $^{2}$Full bandpass.\\
 $^{3}$Count rate after treatment for pile-up. Initial count rate was $689\pm0.1$}

\end{minipage}
\end{table*}

There are 14 confirmed UCXBs with measured orbital periods of less than one hour 
\citep[e.g. see][for a comprehensive list of candidate and confirmed UCXBs]{2012A&A...543A.121V}. 
The majority of these sources has been observed by multiple X-ray observatories.
The present work focuses on {\it XMM-Newton} observations.

After analyzing all {\it XMM} observations of confirmed UCXBs, we have selected five sources for further analysis.
The details of these observations are summarized in Table~\ref{tab:obs}.
The five sources were selected due to their simple, power law dominated spectrum above 2.5\,keV.
Due to calibration uncertainties, as well as features inherent in the source emission, most of the sources
in our sample display increasing spectral complexity below 2-2.5\,keV. Since the focus of our investigation lies
in the high energy part of the spectrum and a detailed  description of the spectral continuum is beyond the scope of this paper, 
we have chosen to ignore all energy channels below 2.5\,keV.
This configuration allows for a more reliable detection of the Fe $\rm K{\rm\alpha}$ line.
The only exception to this treatment is SAX J1808.4-3658
where a thermal component is strongly required by the fit, despite our channel selection.
As a result -- for SAX J1808.4-3658 -- we decided to exclude only energy channels below 1\,keV and include the additional spectral component in our model.
The two LMXBs that constitute our control sample feature similar hard-state spectra.
They have been chosen in order to verify our data analysis procedures in the full range of count rates, spanned by our UCXB sample.

Data reduction was performed using the {\it XMM-Newton} Data Analysis software SAS version 13.5.0. 
The present work focuses on the hard (above 1 keV) part of the spectrum and particularly the presence or absence of iron 
emission features at $\approx 6-7$\,keV. The effective area of EPIC-pn at $\approx 7$\,keV is approximately five times higher than 
that of MOS. Furthermore, during all observations analyzed below, at least one of the MOS detectors was operated in imaging mode. Due to the brightness
of our sources, most of the observations taken in this mode were suffering from severe pile-up.
 On the other hand -- with the exception of 4U 0513-40 -- pn was operating in timing mode during all observations analyzed in this work.
In this mode, photon coordinates are resolved only in one dimension, along the column axis, allowing for high speed CCD read out. 
Apart from offering high time resolution, the timing mode is particularly suited for observing bright sources, since it  allows
for a higher non-piled up count rate.
For these reasons our work is based exclusively on pn data. 


Source photons for all pn observations, taken in timing mode, were extracted 
for RAWX from 25 to 50 and background for RAWX from 3 to 5, where RAWX is the coordinate
along the column axis.
For 4U 0513-40, where pn was operating in imaging mode, 
we extracted the source spectrum from a 27\arcsec circle centered at the source. Background was extracted according to 
latest EPIC calibration notes\footnote{http://xmm2.esac.esa.int/docs/documents/CAL-TN-0018.pdf}
from a source-free region at the same {\small{RAWY}} position as the source region. 
In all cases we extract photons with pixel pattern less than 5. 
All pn observations were checked for pile-up, using the SAS task \texttt{epatplot}. With the exception of
SAX J1808.4-3658 no signatures of pile-up were found for any of the sources in our samples. 
Due to the high count rate of SAX J1808.4-3658 the pn data suffered from pile-up.
To minimize pile-up effects we removed the three central columns from our extraction region.
A subsequent \texttt{epatplot} test
confirms that pile-up effects have been adequately treated.
Lastly, event files for all sources observed in timing mode were treated with \texttt{ epfast}.
This SAS tool is a default setting in SAS 13.5 and corrects possible Charge Transfer Inefficiency (CTI) effects, due to high count rate.

Analysis is done using the {\small {XSPEC}} spectral fitting package, version 12.8.0
\citep{1996ASPC..101...17A}. Prior to analysis, all spectra were re-binned to ensure a minimum of 25 counts per energy channel.
All spectra are fitted with a simple power law, with exception of SAX J1808.4-3658 where an additional
thermal black body component was required.
The iron line was modeled with a Gaussian and when not detected an upper limit on its EW was calculated at 90\% confidence. 
Best-fit parameters for the spectral continuum of all objects along with their distances and
luminosities calculated in the 2.5-10\,keV range are summarized in Table \ref{tab:cont}.
Best-fit parameters and EWs for Fe $\rm K{\rm\alpha}$ emission lines (or upper limits in the case of non detection) are presented in Table \ref{tab:lines}. 

\begin{table*}
 \centering
 \begin{minipage}{157mm}
  \caption{Best-fit parameters for spectral continuum. All uncertainties are $1\,{\rm \sigma}$. }
\label{tab:cont}
  \begin{tabular}  {lccccc}
  \hline
 Sources   &    &   Power Law & &Luminosity&Distance\\
 \cline{1-6}
 \hline                                            
          &            $ {\rm \Gamma}$ &Norm                                        &   $\rm{{\chi_{\nu}}^{2}}$(d.o.f)   &Lx$^{1}$&\\
          &                &($10^{-2}\,{\rm ph\,keV^{-1}\,cm^{-2}\,s^{-1}}$)  &   &($10^{36}{\rm \,erg\,s^{-1}}$)&(kpc)\\
  \hline
  With faint or no Fe $\rm K{\rm\alpha}$ line\\
  \hline
4U 0513-40    & $2.04_{-0.01}^{+0.02}$  & $2.29 _{-0.04}^{+0.09}$& 1.05(1065)&0.9&12.2$^{2}$\\
2S 0918-549   & $2.19\pm0.05$  & $5.93\pm0.05$   &1.01(1470)&0.4&4.8$^{3}$\\
XTE J1807-294 & $2.07 \pm0.02$  & $7.22\pm0.3$   &1.00(1265)&1.1&8.3\\
\hline
  With bright Fe $\rm K{\rm\alpha}$ line\\
  \hline
4U 0614+091        & $2.28\pm0.01$  & $32.91 _{-0.7}^{+0.8}$      &1.01(1490)&0.6& 3.2$^{4}$\\
4U 1916-05   &  $1.84\pm0.02$  & $6.22\pm0.20$  &1.07(1413)&1.6&8.9$^{5}$\\
\hline 
Control sample\\
  \hline
4U 1705-44         &  $1.91\pm0.01$  & $8.85\pm0.16$  &1.05(1493)&1.5&7.4$^{6}$\\
SAX J1808.4-3658   &  $2.19\pm0.01$  & $76.5\pm0.01$  &1.30(1795)&2.0&3.5$^{7}$\\
\hline
\end{tabular}

\medskip

{The models used are: \texttt{phabs$\cdot$powerlaw} (4U 0513-40, 2S 0918-549, XTE J1807-294), 
\texttt{phabs$\cdot$(powerlaw+gaussian)} (4U 0614+091, 4U 1705-44), \texttt{phabs$\cdot$(powerlaw+gaussian)$\cdot$gabs$\cdot$gabs} (4U 1916-05)
and \texttt{phabs$\cdot$(blackbody+powerlaw+gaussian)} (SAX J1808.4-3658). Here \texttt{phabs} is the XSPEC model for photoelectric absorption
and \texttt{gabs} is the XSPEC model for a Gaussian absorption line.
The paramaters of the black body component in SAX J1808.4-3658 are kT=$0.23\pm0.003$ and norm=0.007$\pm0.0002$.

$^{1}$ Calculated in the 2.5-10\,keV range.
$^{2}$ \cite{2003A&A...399..663K}
$^{3}$ \cite{2004MNRAS.354..355J}
$^{4}$ \cite{2010A&A...514A..65K}
$^{5}$ \cite{2008ApJS..179..360G}
$^{6}$ \cite{2008ApJS..179..360G}
$^{7}$ \cite{2009ApJ...694L..21C, 2009A&A...493L..39P}}

\end{minipage}
\end{table*}

\begin{figure}
  \centering
\includegraphics[trim=0cm 1.20cm 0cm 0cm, clip=true, width=0.5\textwidth, angle=-90]{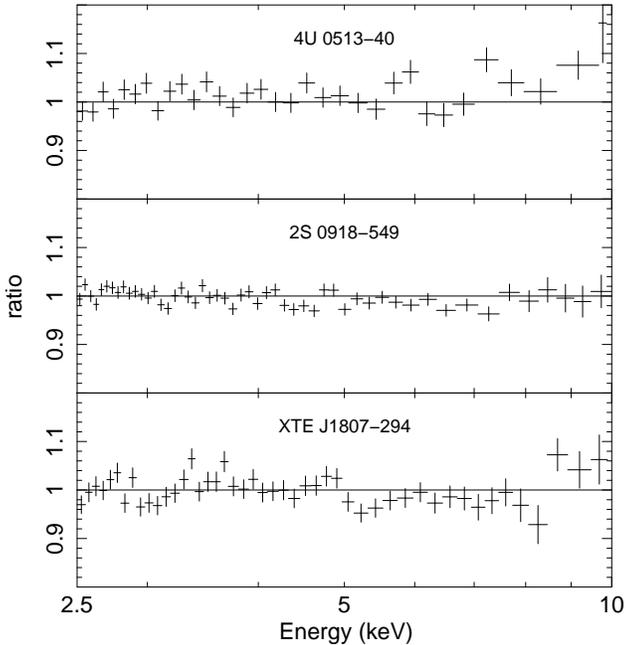}
\caption{Ratio of the data to continuum model vs energy (keV) for three UCXB systems where no Fe line was detected.
Best-fit parameters of the continuum models and upper limits on Fe line flux and EW are listed in Tables~\ref{tab:cont} and~\ref{tab:lines}, respectively.}

\label{fig:noline}
\end{figure}
 
\subsection{4U 0513-40}
4U 0513-40 is located at a distance of $\approx12.2$\,kpc \citep{2003A&A...399..663K} and has an observed orbital period of 17\,min \citep{2011MNRAS.414L..41F}. 
It has been observed by {\it BeppoSAX} , {\it Chandra}, {\it INTEGRAL}, {\it RXTE} and {\it XMM-Newton}.  
{\it XMM-Newton} observations have been analyzed by \cite{2005ApJ...627..926J}, where they report no significant emission or absorption features in the spectrum.
No constraints on the donor composition -- based on optical spectroscopy -- have been reported.

At least nine thermonuclear X-ray bursts have been reported for 4U 0513-40. Two of them have been observed
by Chandra and BeppoSAX \citep[][respectively]{2001ApJ...550L.155H, 2011MNRAS.414L..41F}, and another 7 by 
RXTE \citep{2008ApJS..179..360G}. The latter have been detected after analyzing more than ten years of 
RXTE observations. 

{\it XMM-Newton} observed 4U 0513-40 on 2003 April 1 for 24 ks.  All EPIC-cameras were working 
in imaging mode with MOS1 and MOS2 suffering from considerable pile-up.
The net source count rate is $35.40\pm0.05\,{\rm c/s}$.

The spectrum of 4U 0513-40 above 2.5\,keV, was adequately described by an absorbed power law. 
No Fe $\rm K{\rm\alpha}$ emission line was detected in the spectrum.
An upper limit for the EW of the iron line was calculated by including a Gaussian with a fixed width
of 0.5\,keV, centered between 6.4-6.9\,keV. The source has a moderate luminosity.

\subsection{4U 0614+091}
4U 0614+091 lies close to the galactic plane, at a distance of $\approx3.2$\,kpc \citep{2010A&A...514A..65K} 
and has a measured optical period of $\approx50$\,min 
\citep{2008PASP..120..848S}. It has been observed by {\it ASCA}, {\it BeppoSAX}, {\it Chandra}, 
{\it RXTE}, {\it  Swift} and {\it XMM-Newton}. 
Numerous authors report the detection of a broad emission-like feature near 0.7\,keV 
\citep[e.g.][]{1994ApJ...422..791C, 1999A&A...349L..77P,
2010MNRAS.407L..11M, 2010ApJ...725.2417S}. Nevertheless, using data taken with the  {\it Chandra}
 {\it Low-Energy Transmission Grating Spectrometer} (LETGS),  \cite{2001ApJ...546..338P} failed to detect the feature.
They do, however report an unusually high Ne/O abundance ratio based on absorption edges. Furthermore,
\cite{2001ApJ...560L..59J}, analyzing archival {\it ASCA} data, argue that the broad line-like feature,
reported by previous authors, can be explained by neon overabundance, attributed to neon-rich material local to
the binary. \cite{2010ApJ...725.2417S} analyzing data taken with the {\it High-Energy Transmission Grating Spectrometer} 
(HETGS) on board {\it Chandra}, confirm the existence of
excess optical depth near the Ne K edge. They also detect an extremely broadened O{\small VIII} $\rm Ly{\rm\alpha}$ 
emission feature, which they attribute to emission by highly ionized material in the inner parts of the disk.
In order to account for the line broadening, the authors need to invoke the effects of
gravitational broadening. Fitting a Laor profile  \citep{1991ApJ...376...90L}
to the residuals reveals an inclination angle of the order of $\approx88^{\rm{o}}$.
However, there is no compelling evidence suggesting an edge-on inclination for 4U 0614+091 
and there have been no dips observed in its light curve. To overcome this problem,
\cite{2010MNRAS.407L..11M}, suggest that a similar broad emission feature -- this time detected in
{\it XMM-Newton} RGS observations -- is due to a gravitationally broadened  
O{\small VIII} emission line, that is the result of reflection of the source's primary spectrum off a C/O-rich disk. In this scenario, 
line broadening is partly due to effects of Compton scattering and their fit does not require such 
a high inclination angle. \citeauthor{2010MNRAS.407L..11M} also report the absence of an iron emission line 
in their analysis of MOS2 data of the same observation. They propose that 
this could also be attributed to non solar composition of the accretion disk. Nevertheless \cite{2010A&A...522A..96N},
analyzing pn data of the same {\it XMM-Newton} observation, report a broad emission feature at $\approx6.8\,\rm{keV}$ with an
EW of the order of $\approx160$\,eV. In a more recent publication, \cite{2014arXiv1403.1432M} reanalyze 
RGS observations of 4U 0614+091, this time using a modified version of the {\small{XIILVER}} reflection model \citep{2010ApJ...718..695G, 2013ApJ...768..146G}.
{\small{XIILVER}} is adjusted to mimic a C/O-rich disk by increasing the abundances of carbon and oxygen relative to hydrogen by more than a hundredfold. 
This produces a disk reflection spectrum featuring a strong O{\small{VIII}} emission line which is then used to model RGS observations of 4U 0614+091. 
The authors conclude that the broad positive residuals at $\approx$0.7\,keV are due to a relativistically broadened O{\small{VIII}} emission line 
caused by reflection off a C/O-rich disk.

%


Absence of H or He lines, along with the presence of C{\small{I-IV}} and O{\small{I-III}} emission lines in optical 
spectra of 4U 0614+091, taken with {\it VLT}, suggest a C/O-rich accretion disk and donor star \citep{2004MNRAS.348L...7N, 2006A&A...450..725W}. 
In particular, \citeauthor{2006A&A...450..725W} use synthetic spectra, generated by a non-LTE accretion disk model,
to place an upper limit of 10\% on the abundance of H and/or He present in the accretion disk.
These conclusions seem to be at odds with the fact that 4U 0614+091 is a source 
of multiple X-ray bursts, with a measured recurrence time ranging from once every few weeks to once every $\approx10\,$days 
\citep{2010A&A...514A..65K, 2012ApJ...760..133L}. Namely, \citeauthor{2010A&A...514A..65K} --  
based on archival data collected from multi-instrument observations -- calculate an average burst recurrence rate of about once every one to
two weeks. \citeauthor{2012ApJ...760..133L} use data collected with 
the Gamma-ray Burst Monitor (GBM) aboard the {\it Fermi Gamma-ray Space Telescope} to calculate 
a burst recurrence time of $\approx12$\,d, with the closest burst pair recorded within 2.8\,d. Furthermore, by creating ignition
models for type I X-ray bursts, following the work of \cite{2001ApJ...559L.127C}, \citeauthor{2010A&A...514A..65K} 
demonstrate that a H or He amount, substantially larger than the inferred 
upper limit of 10\%, is required to simulate the characteristics of the observed bursts.

{\it XMM-Newton} observed 4U 0614+091 on 2001 March 13. There were two subsequent observations
for $\approx11$\,ks and  $\approx17$\,ks. The MOS1, MOS2 and pn detectors
were only active during the second observation, therefore we focus on the second observation only. 
The net source count rate is $252.1\pm0.17\,{\rm c/s}$.

Modeling 4U 0614-091 with an absorbed power law leaves increased positive residuals
in the 6-7\,keV range (Fig.~\ref{fig:lines}). We account for these by adding a Gaussian to our initial model. 
This reveals a bright and broad Fe $\rm K{\rm\alpha}$ emission line and improves our fit by a $\Delta\chi^2$ of
89, which corresponds to a more than 4$\sigma$ significance for 3 d.o.f..
The line is centered at $\approx6.64$\,keV with an EW of $\approx111$\,eV
and a width of  $\approx0.7$\,keV.

\subsection{2S 0918-549}
2S 0918-549 has a measured orbital period of 17.4\,min \citep{2011ApJ...729....8Z} and lies at a distance of $\approx4.8$\,kpc \citep{2004MNRAS.354..355J}. It has been observed
by {\it ASCA}, {\it BeppoSAX}, {\it Chandra}, {\it RXTE}, {\it XMM-Newton} and {\it VLT}. \cite{2001ApJ...560L..59J} 
analyzed archival {\it ASCA}  observations of 2S 0918-549. 
They suggested an O/Ne chemical composition of the accreting material based on an unusual Ne/O number ratio compared to what is 
expected for the interstellar medium. This non-solar relative abundance of Ne was attributed to enrichment of the local ISM by the donor material.
However further examination \citep{2003ApJ...599..498J, 2005ApJ...627..926J}
revealed that the Ne/O ratio varies between different  {\it Chandra}, {\it XMM-Newton} and
{\it ASCA} observations of 2S 0918-549. This variability, combined with a tenfold decrease in flux 
between the  {\it Chandra} and {\it XMM-Newton} observations and the earlier one by {\it ASCA},
led to the conclusion that the unusual Ne/O ratio maybe due to ionization effects and is not proof of the donor composition. 
In their analysis of 2S 0918-549, \cite{2003ApJ...599..498J} do not comment on the existence or absence of the iron  $\rm K{\rm\alpha}$ emission line.
{\it VLT} data, analyzed by \cite{2004MNRAS.348L...7N} tentatively suggest a C/O-rich chemical composition of the accretion disk and donor star.

2S 0918-549 is also known to produce sporadic X-ray bursts. At least six type I X-ray bursts have been reported
for this binary between 1996 and 2004. Two have been observed by \cite{2001ApJ...553..335J} and \cite{2002A&A...392..885C} and another 
four by \cite{2005A&A...441..675I}. \citeauthor{2005A&A...441..675I} in contrast to what is suggested by optical data, 
suggest the possibility of a He-rich donor, based on the system's bursting activity.

{\it XMM-Newton} observed 2S 0918-549  on 2001 May 5 for 40 ks. MOS1 and pn cameras were operated in timing mode, while the MOS2 camera was operated in imaging mode.
The net source count rate is $60.47\pm0.06\,{\rm c/s}$.

Similarly to 4U 0513-40, 2S 0918-549 has a moderate luminosity and is adequately modeled with an absorbed power law. No emission feature was required by the fit.

\subsection{XTE J1807-294}
XTE J1807-294 is a transient, accreting  millisecond pulsar.  
\cite{2003ATel..127....1M}, using {\it Chandra} observations, detected the source in the direction of the galactic bulge, 
suggesting a distance of  $\approx8.3$\,kpc. X-ray pulsations observed
using {\it RXTE}  give an orbital period of $\approx 17$\,min \citep{2003ATel..127....1M}. {\it XMM-Newton} observations
were analyzed by \cite{2003ApJ...594L..39C}. No emission or absorption features were detected, and an upper limit of 18-25\,eV is placed on 
the EW of the iron line. Similarly, simultaneous {\it INTEGRAL}, {\it XMM-Newton} and {\it RXTE}  observations by \cite{2005A&A...436..647F} 
revealed no evidence of emission or absorption features in the source continuum.  
{\it VLT} observations by \cite{2009A&A...508..297D} failed to detected the source's optical counterpart.
No type I X-ray bursts have been reported for this system.

{\it XMM-Newton} observed XTE J1807-294 on 2003 March 22 for $\approx 17$\,ks during an outburst. 
MOS1 and MOS2 were operating on imaging mode and both suffered from pile-up.
The net source count rate is $35.42\pm0.11\,{\rm c/s}$.

Again, the spectrum was modeled with an absorbed power law and no Fe $\rm K{\rm\alpha}$ emission line was detected.
The data-to-model ratio vs energy plot for XTE J1807-294 is presented in Fig.~ \ref{fig:noline} along with those of
4U 0513-40, 2S 0918-549 for which also no iron line was detected.

\subsection{4U 1916-05}
4U 1916-05 is a dipping source (its light curve exhibits periodic intensity dips), 
with an orbital period of 50\,min \citep{1982ApJ...253L..67W} and a distance of 
$\approx8.9$\,kpc  \citep{2008ApJS..179..360G}. The characteristics of the source's dipping behavior 
indicate that it is viewed at a large inclination angle, with $\rm i\ge60{}^\circ$ \citep{1988MNRAS.232..647S}. 
{\it ASCA} observations of 4U 1916-05 were analyzed by \cite{2000ApJS..131..571A}. They report a broad emission feature at 5.9\,keV with an
EW of $\approx$87\,eV. The presence of an emission feature in the unusual energy of 6.0\,keV is also reported by \cite{2004A&A...418.1061B} in their
analysis of the {\it XMM-Newton} observation of 4U 1916-05. However this feature was not further investigated as it was not the focus of their paper.
On the other hand, they report the detection of two narrow absorption lines at 6.65 and 6.95\,keV that are consistent with resonant absorption
from Fe{\small XXV} and Fe{\small XXVI} ions, respectively. This detection is repeated in {\it Chandra} observations analyzed by \cite{2006ApJ...646..493J}.
The authors also report the existence of narrow absorption lines, due to the presence of H-like neon, magnesium, silicon, and sulfur. In addition to X-rays,
optical observations by {\it VLT}, found prominent lines from He{\small I}, He{\small II}, N{\small II} and N{\small III},
consistent with a He-rich accretion disk \citep{2006MNRAS.370..255N}. 

4U 1916-05 is also a known X-ray burster. It exhibits short bursts that are consistent with moderate to high accretion rates, and have
a recurrence time of $\approx$6.2\,hr \citep{2008ApJS..179..360G}. 

{\it XMM-Newton} observations of 4U 1916-05, analyzed in this work, were performed on September 25, 2002 for 17 ks. All
EPIC cameras were operating in timing mode. The net source count rate is $55.47\pm0.11\,{\rm c/s}$.

\begin{figure}
\centering
\includegraphics[trim=0cm 1.20cm 0cm 0cm, clip=true, width=0.5\textwidth, angle=-90]{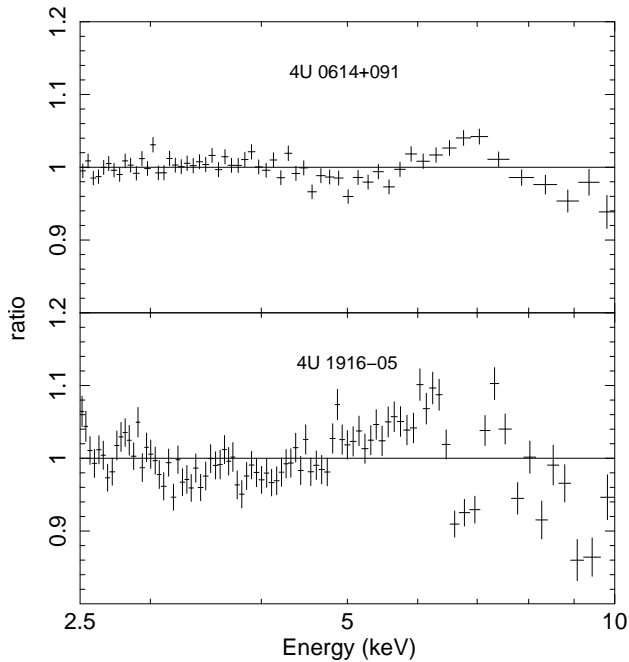}
\caption{Ratio of the data to continuum model vs energy for 4U 1916-05 and 4U 0614-091 whose spectra feature a bright iron line. 
Best-fit parameters of the continuum models are listed in Table~\ref{tab:cont}. Fe line flux and EW for Gaussian
are listed in Table~\ref{tab:lines}.}
\label{fig:lines}
\end{figure}

Fitting the spectrum of 4U 1916-05 with an absorbed power law revealed a complex structure 
with both positive and negative residuals between $\approx6-7$\,keV (Fig.~\ref{fig:lines}). 
As has been suggested before \citep[e.g.][]{2004A&A...418.1061B}, the negative residuals could be interpreted as 
absorption lines at $\approx6.65$ and $\approx6.95$\,keV, due to resonant absorption by Fe{\small XXV} and Fe{\small XXVI} ions.
To account for these features, we add two narrow Gaussian absorption lines to our model. The first one -- corresponding to absorption due to He-like iron --
is centered at $\approx6.66$\,keV and improves our fit by a $\Delta\chi^2$ of 40 for 3 d.o.f..
The second -- corresponding to absorption due to H-like iron --
is centered at $\approx6.92$\,keV and reduces the $\chi^2$ value by 20 for 3 d.o.f..
The two absorption lines  are strongly required by the fit and are of high (more than 3$\sigma$) significance.
They can be attributed to highly ionized plasma in the vicinity of the accretion disk.

The remaining positive residuals suggest a complex broad emission feature in the 5-7\,keV range.
Fitting this feature with a Gaussian reveals a broad emission 
``line'' at the unexpected energy of $\approx5.46$\,keV with a width of $\approx1.64$\,keV. 
The EW of this feature is $\approx595$\,eV. Adding the Gaussian line to our model,
 improves our fit massively,  by a $\Delta\chi^2$  of 187 for 3 d.o.f..
However, the unusual values of the line parameters and the fact that this is a dipping source, 
could indicate that this feature is an artifact resulting from our extraction of a 
single spectrum for the entire observation, during both dipping and persistent phases.
To investigate this possibility, we extracted two additional spectral sets. One taken from events that 
were recorded only during the persistent phase and a second one during the dipping phase.
The results are qualitatively identical to the ones obtained for the full observation.
All three spectra (dipping, persistent and combined) show strong evidence of a broad emission feature 
located at $\approx5.5-6$\,keV.
Obviously this feature is too complex to be fitted with a simple Gaussian.
Hence the parameters obtained have no physical meaning. 
Nevertheless, in the framework of our model, the statistical significance of a strong iron emission line
in the 6-7\,keV range is beyond doubt.
Further examination of the detailed spectral shape of the emission line is beyond the scope of this paper.
If we fix the line energy at 6.4\,keV and the line width at 0.5\,keV
we obtain an emission line with an EW of $\approx95$\,eV with more than $4{\rm\sigma}$ significance.

\subsection{Control Sample}

\subsubsection{4U 1705-44}
4U 1705-44 is a typical LMXB containing a neutron star accretor with a weak magnetic field.
Its distance is measured at $\approx 7.4$\,kpc \citep{2008ApJS..179..360G}. The source exhibits frequent type I
X-ray bursts \citep[e.g.][]{1989ApJ...339.1044G} and has been observed during hard and soft states. In both states its 
spectrum features a bright broad iron $\rm K{\rm\alpha}$ emission line
\citep[e.g][]{2005ApJ...623L.121D, 2009MNRAS.398.2022D, 2010ApJ...720..205C, 2013A&A...550A...5E}. 
Multiple observations of 4U 1705-44 have been performed by all major X-ray telescopes. 4U 1705-44 is a frequent burster 
with a recurrence time of $\approx0.91$\,hr \citep{2008ApJS..179..360G}.

In the present work we reanalyze {\it XMM-Newton} observations performed on September 26, 2006 during a hard state.
Only pn and RGS were active during the observation with pn operating in timing mode. The net source count rate is $27.23\pm0.05\,{\rm c/s}$.

As expected for the LMXBs in our control sample, we detect
strong positive residuals in the 6-7\,keV range (Fig. \ref{fig:control}), characteristic of a strong iron line.
Fitting the emission feature  with a Gaussian yields an emission line with an EW of $\approx52$\,eV (for details see Table~\ref{tab:lines}).

 \subsubsection{SAX J1808.4-3658}
SAX J1808.4-3658 is a transient LMXB with an orbital period of $\approx2.01$\,hrs \citep{1998Natur.394..346C}.
Its distance is measured at $\approx 3.5$\,kpc \citep{2009ApJ...694L..21C, 2009A&A...493L..39P}.
It was the first millisecond pulsar discovered \citep{1998Natur.394..344W}. The source's quiescent periods are 
interrupted by short outbursts approximately every 2.5 years. SAX J1808.4-3658 has been observed by 
{\it ASCA}, {\it BeppoSAX}, {\it Chandra}, {\it INTEGRAL}, {\it  Swift}, {\it RXTE} and  {\it XMM-Newton}
during both quiescent and bursting periods. {\it VLT} observations have also been performed during its 2008 September-October
outburst. Analysis of source spectra during outbursts, have revealed the existence of a Fe $\rm K{\rm\alpha}$ emission line
\citep[e.g][]{1998A&A...338L..83G, 2009ApJ...694L..21C, 2009A&A...493L..39P}. Lastly, the source is a
burster with a recurrence time of  $\approx21$\,hr \citep{2008ApJS..179..360G}.

We reanalyze  the 63\,ks  {\it XMM-Newton} observation of October 1st 2008 during outburst. The MOS1 camera operated in imaging mode
while MOS2 and pn in timing mode. 
The net source count rate is $300.3\pm0.09\,{\rm c/s}$.
 
\begin{figure}
  \centering
\includegraphics[trim=0cm 1.20cm 0cm 0cm, clip=true, width=0.5\textwidth, angle=-90]{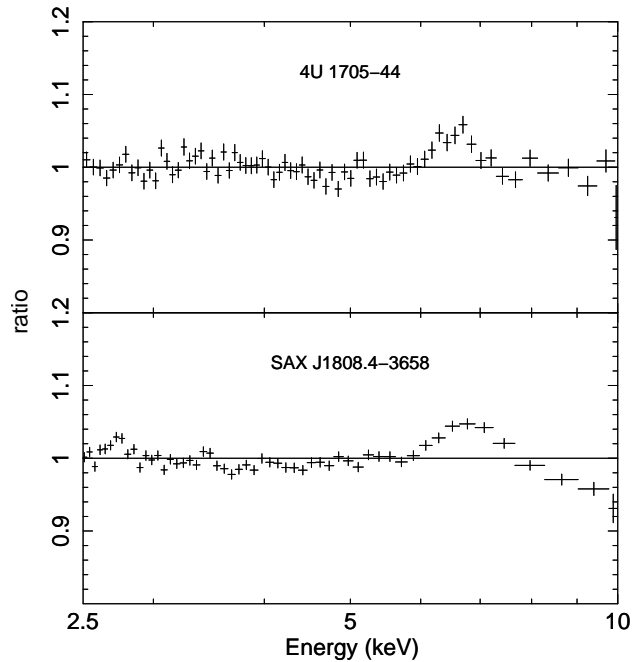}
\caption{Ratio of the data to continuum model vs energy for the two LMXBs in our control sample.
Best-fit parameters of the continuum models are listed in Table~\ref{tab:cont}.
Fe line flux and EWs are listed in Table~\ref{tab:lines}.}
\label{fig:control}
\end{figure}

SAX J1808.4-3658 is fitted using a black body and a power law spectrum.
Examination of the data-to-model ratio of our fit, reveals a strong emission feature at the 6-7\,keV range (Fig. \ref{fig:control}). Fitting this 
feature with a Gaussian reveals a broad emission line centered at $\approx6.61$\,keV with a width of $\approx1.13$\,keV
and an EW of $\approx246$\,eV. The emission feature is characteristic of the iron $\rm K{\rm\alpha}$ emission line 
expected in X-ray reflection spectra originating in typical LMXBs with main sequence or red giant donors. 
The broadness of the line obviously justifies further inquiry on different possible mechanisms
that would explain it. However, such a task has already been accomplished by \cite{2009A&A...493L..39P} and is not the focus of the present work.

\begin{table*}
 \centering
 \begin{minipage}{143mm}
  \caption{Best-fit parameters of iron $\rm K{\rm\alpha}$ emission line. 
  In case of no detection an upper limit at $90\%$ confidence level is given. All errors are 1$\rm \sigma$.}
\label{tab:lines}
  \begin{tabular}  {lccccl}
  \hline
  
 Sources &  & Gaussian&  &&Bursts\\
 \cline{1-6}
 \hline 
        &Norm             &  LineE      & ${\rm \sigma}$     &EW     &Rec. Time$^{1}$\\
        & ($10^{-5}\,{\rm ph\,cm^{-2}\,s^{-1}}$)     &  (keV)      & (keV)              &   (eV)&\\
  \hline
  With faint or no Fe $\rm K{\rm\alpha}$ line\\
  \hline 
4U 0513-40        & $<0.92$ & 6.4-6.9& 0.5$^{2}$&$<20$&9 bursts\\
2S 0918-549       & $<0.63$ & 6.4-6.9& 0.5$^{2}$&$<7 $&6 bursts\\
XTE J1807-294     & $<1.25$ & 6.4-6.9& 0.5$^{2}$&$<10$&0 bursts\\
  \hline
  With Fe $\rm K{\rm\alpha}$ line\\
  \hline
4U 0614+091        &$32.9_{-7.4}^{+7.9}$  & $6.64\pm0.08$  & 0.67$_{-0.08}^{+0.10}$&$111.1_{-17}^{+20}$&$\approx$10\,d\\
4U 1916-05   &$27.7\pm2.47$&6.4$^{2}$&0.5$^{2}$  &$95\pm12$&$\approx$6.2\,hr\\
  \hline
Control sample\\
  \hline
4U 1705-44        & $13.2_{-1.82}^{+1.95}$ & $6.53\pm0.05$&$0.34\pm0.05$&$51.7_{-12}^{+13}$&$\approx$0.91\,hr\\

SAX J1808.4-3658  & $173\pm27$ & $6.40\pm0.07$&$0.83_{-0.11}^{+0.10}$&$131\pm20$&$\approx$21\,hr\\
\hline
\end{tabular}
\medskip

{$^{1}$ Recurrence time when available. Otherwise total number of recorded bursts. All references regarding bursting activity are given in Section 2.

$^{2}$ The parameter was fixed.}

\end{minipage}
\end{table*}

\section{Discusssion}
We have analyzed the spectra of 5 confirmed UCXBs with H-poor donors and low to moderate luminosities. Specifically, we investigated the existence of a  
Fe $\rm K{\rm\alpha}$ line in their spectra. Three objects in our sample -- 
namely 2S 0918-549, XTE J1807-294, 4U 0513-40 -- display no obvious emission features in the energy range
between 6 and 7\,keV. This result is in agreement with the works of previous authors \citep[e.g.][]{2003ApJ...594L..39C, 2005ApJ...627..926J}.
On the other hand, systems 4U 0614+091 and 4U 1916-05 display a bright iron $\rm K{\rm\alpha}$ line. This emission feature is broad and more complex than
a simple Guassian would describe, but there is no doubt that strong iron emission is evident in both sources.
The spectra of 4U 0614+091 and 4U 1916-05 are similar (e.g. see the ratio plots in figures~\ref{fig:lines} and~\ref{fig:control})
to the spectra of typical LMXBs like 4U 1705-44 and SAX J1808.4-3658, also analyzed in this work. 

\begin{figure}
  \centering
\includegraphics[trim=0.2cm 0cm 0cm 0cm, clip=true, width=0.52\textwidth, angle=0]{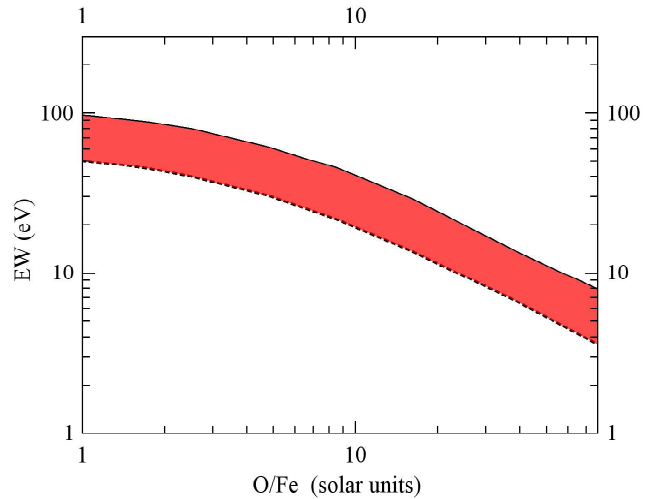}
\caption{ Dependence of Fe $\rm K{\rm\alpha}$ line EW (eV) on O/Fe ratio, given in units of its solar value.
The curves were calculated for an incident spectrum with power law spectral shape with an index of 2.2. The solid curve is calculated for
a face-on configuration ($0^{{\rm o}}$ inclination) and the dotted line for an $\approx$edge-on configuration ($80^{{\rm o}}$ inclination).}
\label{fig:code}
\end{figure}

\subsection{X-ray diagnostics}
In Koliopanos et al. (2013) we showed that non-solar composition of the accretion disks in UCXBs can have a powerful effect on
fluorescent emission lines appearing in their spectra. Namely, the bright iron $\rm K{\rm\alpha}$ line, typically found in
spectra of LMXBs with main sequence or red giant donors, is expected to be severely suppressed in the case of UCXBs with  C/O or O/Ne/Mg WD donors.
This is due to screening by oxygen in the C/O or O/Ne dominated material.
On the other hand, due to the lower ionization threshold of He,
the iron line in spectra of UCXBs with He star or He WD donors, is expected to retain its nominal strength, mostly determined by iron abundance. 
Ionization of the disk material at high mass accretion rates may lead to luminosity dependence of this behavior.
Specifically, line suppression due to screening by oxygen, is expected to 
take effect in objects with moderate luminosities (${\rm L}\rm_X\lesssim$ a few $10^{37}\rm erg\,s^{-1}$).

\subsubsection{Sources with no Fe $\rm K{\rm\alpha}$ emission line}
Using results of our Monte Carlo simulation from Koliopanos et al. (2013)
we can place constraints on the O/Fe ratio of the accretion disk in the three sources 
for which an iron line was not detected. The code simulates reflection off a homogeneous slab of infinite optical depth 
that is comprised of cold and neutral material . For this example, the primary, incident radiation has a power law shaped spectrum with a
spectral index of 2.2 and is emitted isotropically from a point source above the disk surface. 
In order to study the whole range of possible configurations,
from face-on to $\approx$edge-on view, we run two sets of calculations, where the reflection spectrum is registered
at two distinct viewing angles of $0^{\rm o}$ and $80^{\rm o}$ respectively. 
Since there are no dips in the light curves of the three systems, we do not investigate a viewing angle larger than $80^{\rm o}$. 
The code simulates fluorescence $\rm K{\rm\alpha}$ and $\rm K{\rm\beta}$ lines for elements from Z=3 to 30 
and reports their EWs with respect to the total emission, which is a mixture of both the primary and reflected emission 
(for details see Appendix A in Koliopanos et al. 2013).

We generate a grid of EW values of the Fe $\rm K{\rm\alpha}$ fluorescence line at 6.4\,keV, calculated for a sequence of
increasing C/O abundance, starting with solar-like material. Abundances of H and He are reduced along the sequence, thus conserving the total number of nucleons. 
Similarly to Koliopanos et al. mass fractions of all other elements remain fixed at their solar values, as well as the abundance ratio of carbon
and oxygen. The position along this sequence is given in terms of O/Fe ratio, in units of its solar value.
Obviously, as H and He are ``converted'' to C and O, O/Fe ratio will increase. The maximum value of the O/Fe ratio that
corresponds to a C/O-rich disk -- in which all hydrogen and helium has been replaced by carbon and oxygen -- is $\approx77$ times its solar value. 
Solar abundances for elements with Z=1-30 were adopted from \cite{1992PhyS...46..202F}, elements not listed in
this tabulation were taken from \cite{1998SSRv...85..161G}.
The resulting dependencies of EW on the O/Fe ratio are presented in Fig.~\ref{fig:code}. 

As is evident from the plot, even in the marginal case of a system viewed $\approx$edge-on,
an EW value of 20\,eV -- the highest upper limit measured for the three sources -- corresponds to an O/Fe 
ratio that is more than $\approx$10 times higher than the solar value.  
The EW upper limits of 7 and 10\,eV -- for 2S 0918-549 and XTE J1807-294 -- correspond to  O/Fe 
ratios exceeding $\approx$37 and $\approx$20 times the solar value, respectively. Since there are no indications (e.g. a dipping light curve)
of a large inclination angle for these three sources, the real lower limits are probably higher. These results
place a strong constraint on the chemical composition of these systems. 
Namely, we conclude that the lack of an iron emission line in the spectra of 2S 0918-549, XTE J1807-294 and 4U 0513-40 
is the result of a C/O or O/Ne/Mg-rich accretion disk and donor star in these systems.

A final point of interest regarding  Fig.~\ref{fig:code} is the fact that in this example, our calculations predict
a maximum Fe $\rm K{\rm\alpha}$ EW value of $\approx$100\,eV. This value is in agreement with the values 
obtained by the fits of 4U 0614+091 and 4U 1705-44 but it is relatively smaller than the value obtained for SAX J1808.4-3658
and significantly smaller than the one obtained for 4U 1916-05. 
This apparent discrepancy is mostly due to the fact that a simple Gaussian profile was used to fit broad emission features that 
have a more complicated shape than a Gaussian can describe. This is particularly evident in 4U 1916+091
which is a confirmed edge-on viewed system with a dipping light curve. Implementation of more sophisticated models
such a Laor profile or XSPEC model \texttt{diskline} \citep{1989MNRAS.238..729F} yields smaller, 
more realistic EW values in the $\approx100$\,eV range. 
Additionally, alternative modeling of the underlying spectral continuum -- especially when the full energy range is considered --
can result in different values for the EW of the line. For instance a different choice of the spectral continuum 
of SAX J1808.4-3658 in \cite{2010A&A...522A..96N} resulted in an EW of  $\approx$30\,eV for the iron line. A value much smaller than 
the one derived in this work ($\approx130$\,eV) and in that of \cite{2009A&A...493L..39P} ($\approx120$\,eV). 
However, a detailed investigation of the emission line profiles and the shape of the spectral continuum was beyond the scope of this work, where we are only
interested in the fact of the existence of the line.
Therefore, energies below 2.5\,kev were ignored and an absorbed power law and a simple Gaussian model were used to model the spectral continuum and the emission lines.

\subsubsection{Sources with Fe $\rm K{\rm\alpha}$ emission line}
The  presence of a strong Fe $\rm K{\rm\alpha}$ line in the spectra of two remaining sources in our sample, 4U 1916-05 and 4U 0614+091
is an indication of a He-rich accretion disk and donor star in these systems. Namely, according to our model, the presence of a strong iron line, 
in the spectrum of a moderately luminous object, requires an O/Fe ratio that is close to the solar value. 
In the context of UCXBs, this would point towards a He-rich donor.

\subsection{Optical spectra and X-ray bursts}
Out of the three systems for which no emission line was detected, only 2S 0918-549 has been studied by means of optical spectroscopy.
The analysis of \cite{2004MNRAS.348L...7N} tentatively suggests a C/O-rich rich donor. This is in agreement with our own conclusions. 
On the other hand, systems 2S 0918-549 and 4U 0513-40 have exhibited sporadic bursting activity during the previous decades 
(see Table~\ref{tab:lines} and relevant subsections).
The characteristics of some of the bursts are compatible with helium burning, a fact that led \cite{2005A&A...441..675I} to suggest
the possibility of a He-rich donor in 2S 0918-549. Nevertheless, over the span of $\approx10$years,
just a handful of bursts has been reported for these two systems \citep[e.g.][]{2005A&A...441..675I, 2008ApJS..179..360G}
and therefore they cannot provide definitive proof of donor composition. It is certainly plausible that small amounts of H and/or He in an
otherwise C/O-rich disk could fuel sporadic bursts.

The strong Fe $\rm K{\rm\alpha}$ line in the spectrum of 4U 1916-05, which is an UCXB system with an optically confirmed He-rich donor \citep{2006MNRAS.370..255N}
is also  in complete agreement with the theoretical predictions presented in Koliopanos et al.. Furthermore, the system's frequent bursting activity
reinforces the arguments in favor of a He-rich donor.
Perhaps more interestingly, however, our argument in favor of a He-rich donor star in 4U 0614+091 is at odds with 
the fact that the system has been classified as C/O rich, based on optical spectroscopy  \citep{2004MNRAS.348L...7N, 2006A&A...450..725W}. 
On the other hand, our estimation readily explains the source's repeated bursting activity as reported by \cite{2010A&A...514A..65K} and \cite{2012ApJ...760..133L} 
and is in complete agreement with the conclusions reached by these authors.

Nevertheless, the lack of He lines in the optical spectrum of 4U 0614+091 \citep{2004MNRAS.348L...7N, 2006A&A...450..725W} and particularly the
upper limit of 10\% He abundance in the disk material, placed by \citeauthor{2006A&A...450..725W} are strong arguments in favor of a C/O rich donor. 
This upper limit is calculated for a disk ${\rm T_{\rm eff}}$ of 28000\,K at a distance of 2000 stellar radii. For this
temperature the dominant ionization stage of helium at $\tau=1$ is He{\small II} \citep[][Figure 6]{2006A&A...450..725W}.
Therefore, according to the \citeauthor{2006A&A...450..725W} model, an accretion disk with a helium abundance of more than 10\%,
would produce stronger He emission lines than the observed upper limits.
However, their model ignores disk illumination and authors acknowledge that
introducing disk irradiation from the NS surface and the comptonizing corona would affect the ionization structure of the disk. 
Namely, if ${\rm T_{\rm eff}}$ is high enough, helium could be fully ionized even in the outer parts of the disk. Such a scenario would promptly
reconcile the lack of He-lines in the optical spectrum, the system's bursting activity 
and the presence of a strong iron line in the reflection spectrum.

\section{Summary and conclusions}
We searched for the iron ${\rm K\alpha}$ in the spectra of five UCXBs with H-deficient donors.
Based on the non-detection of a Fe line and the predictions of \cite{2013MNRAS.432.1264K},
we have concluded that the accretion disk material of three of the objects in our sample (2S 0918-549, XTE J1807-294 and 4U 0513-40) 
has an O/Fe ratio that is at least $\approx$10\,times higher than the solar value. In the context of UCXBs this  
suggests a C/O or O/Ne/Mg-rich donor. 
Furthermore, the presence of a strong Fe ${\rm K\alpha}$ line in the spectra of the 
remaining two systems (4U 0614+091 and 4U 1916-05) indicates a He-rich donor. 
In the case of 2S 0918-549 and 4U 1916-05 our findings are also supported by results obtained through optical spectroscopy.
On the other hand, our suggestion of a He-rich donor in 4U 0614+091 contradicts arguments in favor of a C/O-rich donor,
but is consistent with the source's regular bursting activity.

\section*{Acknowledgements}
The authors would like to thank Dr. Mikhail Revnivtsev for crucial comments and valuable advice.
We also thank Pierre Maggi for frequent helpful discussions. 
This work was supported by the National Science Foundation under
grants PHY 11-25915 and AST 11-09174.

\bibliographystyle{mn2e}
\bibliography{General}

\begin{thebibliography}{70}
\expandafter\ifx\csname natexlab\endcsname\relax\def\natexlab#1{#1}\fi

\bibitem[{{Arnaud}(1996)}]{1996ASPC..101...17A}
{Arnaud} K.~A., 1996, in Astronomical Society of the Pacific Conference Series,
  Vol. 101, Astronomical Data Analysis Software and Systems V, {Jacoby} G.~H.,
  {Barnes} J., eds., p.~17

\bibitem[{{Asai} {et~al}\mbox{.}(2000){Asai}, {Dotani}, {Nagase}, \&
  {Mitsuda}}]{2000ApJS..131..571A}
{Asai} K., {Dotani} T., {Nagase} F., {Mitsuda} K., 2000, \apjs, 131, 571

\bibitem[{{Bildsten} \& {Deloye}(2004)}]{2004ApJ...607L.119B}
{Bildsten} L., {Deloye} C.~J., 2004, \apjl, 607, L119

\bibitem[{{Boirin} {et~al}\mbox{.}(2004){Boirin}, {Parmar}, {Barret},
  {Paltani}, \& {Grindlay}}]{2004A&A...418.1061B}
{Boirin} L., {Parmar} A.~N., {Barret} D., {Paltani} S., {Grindlay} J.~E., 2004,
  \aap, 418, 1061

\bibitem[{{Cackett} {et~al}\mbox{.}(2009){Cackett}, {Altamirano}, {Patruno},
  {Miller}, {Reynolds}, {Linares}, \& {Wijnands}}]{2009ApJ...694L..21C}
{Cackett} E.~M., {Altamirano} D., {Patruno} A., {Miller} J.~M., {Reynolds} M.,
  {Linares} M., {Wijnands} R., 2009, \apjl, 694, L21

\bibitem[{{Cackett} {et~al}\mbox{.}(2010){Cackett}, {Miller}, {Ballantyne},
  {Barret}, {Bhattacharyya}, {Boutelier}, {Miller}, {Strohmayer}, \&
  {Wijnands}}]{2010ApJ...720..205C}
{Cackett} E.~M. {et~al.}, 2010, \apj, 720, 205

\bibitem[{{Campana} {et~al}\mbox{.}(2003){Campana}, {Ravasio}, {Israel},
  {Mangano}, \& {Belloni}}]{2003ApJ...594L..39C}
{Campana} S., {Ravasio} M., {Israel} G.~L., {Mangano} V., {Belloni} T., 2003,
  \apjl, 594, L39

\bibitem[{{Chakrabarty} \& {Morgan}(1998)}]{1998Natur.394..346C}
{Chakrabarty} D., {Morgan} E.~H., 1998, \nat, 394, 346

\bibitem[{{Christian}, {White} \& {Swank}(1994){Christian}, {White}, \&
  {Swank}}]{1994ApJ...422..791C}
{Christian} D.~J., {White} N.~E., {Swank} J.~H., 1994, \apj, 422, 791

\bibitem[{{Cornelisse} {et~al}\mbox{.}(2002){Cornelisse}, {Verbunt}, {in't
  Zand}, {Kuulkers}, {Heise}, {Remillard}, {Cocchi}, {Natalucci}, {Bazzano}, \&
  {Ubertini}}]{2002A&A...392..885C}
{Cornelisse} R. {et~al.}, 2002, \aap, 392, 885

\bibitem[{{Cumming} \& {Bildsten}(2001)}]{2001ApJ...559L.127C}
{Cumming} A., {Bildsten} L., 2001, \apjl, 559, L127

\bibitem[{{D'Avanzo} {et~al}\mbox{.}(2009){D'Avanzo}, {Campana}, {Casares},
  {Covino}, {Israel}, \& {Stella}}]{2009A&A...508..297D}
{D'Avanzo} P., {Campana} S., {Casares} J., {Covino} S., {Israel} G.~L.,
  {Stella} L., 2009, \aap, 508, 297

\bibitem[{{Deloye} \& {Bildsten}(2003)}]{2003ApJ...598.1217D}
{Deloye} C.~J., {Bildsten} L., 2003, \apj, 598, 1217

\bibitem[{{Deloye}, {Bildsten} \& {Nelemans}(2005){Deloye}, {Bildsten}, \&
  {Nelemans}}]{2005ApJ...624..934D}
{Deloye} C.~J., {Bildsten} L., {Nelemans} G., 2005, \apj, 624, 934

\bibitem[{{Di Salvo} {et~al}\mbox{.}(2009){Di Salvo}, {D'A{\'{\i}}}, {Iaria},
  {Burderi}, {Dov{\v c}iak}, {Karas}, {Matt}, {Papitto}, {Piraino}, {Riggio},
  {Robba}, \& {Santangelo}}]{2009MNRAS.398.2022D}
{Di Salvo} T. {et~al.}, 2009, \mnras, 398, 2022

\bibitem[{{Di Salvo} {et~al}\mbox{.}(2005){Di Salvo}, {Iaria}, {M{\'e}ndez},
  {Burderi}, {Lavagetto}, {Robba}, {Stella}, \& {van der
  Klis}}]{2005ApJ...623L.121D}
{Di Salvo} T., {Iaria} R., {M{\'e}ndez} M., {Burderi} L., {Lavagetto} G.,
  {Robba} N.~R., {Stella} L., {van der Klis} M., 2005, \apjl, 623, L121

\bibitem[{{Egron} {et~al}\mbox{.}(2013){Egron}, {Di Salvo}, {Motta}, {Burderi},
  {Papitto}, {Duro}, {D'A{\`i}}, {Riggio}, {Belloni}, {Iaria}, {Robba},
  {Piraino}, \& {Santangelo}}]{2013A&A...550A...5E}
{Egron} E. {et~al.}, 2013, \aap, 550, A5

\bibitem[{{Fabian} {et~al}\mbox{.}(1989){Fabian}, {Rees}, {Stella}, \&
  {White}}]{1989MNRAS.238..729F}
{Fabian} A.~C., {Rees} M.~J., {Stella} L., {White} N.~E., 1989, \mnras, 238,
  729

\bibitem[{{Falanga} {et~al}\mbox{.}(2005){Falanga}, {Bonnet-Bidaud},
  {Poutanen}, {Farinelli}, {Martocchia}, {Goldoni}, {Qu}, {Kuiper}, \&
  {Goldwurm}}]{2005A&A...436..647F}
{Falanga} M. {et~al.}, 2005, \aap, 436, 647

\bibitem[{{Feldman}(1992)}]{1992PhyS...46..202F}
{Feldman} U., 1992, \physscr, 46, 202

\bibitem[{{Fiocchi} {et~al}\mbox{.}(2011){Fiocchi}, {Bazzano}, {Natalucci},
  {Landi}, \& {Ubertini}}]{2011MNRAS.414L..41F}
{Fiocchi} M., {Bazzano} A., {Natalucci} L., {Landi} R., {Ubertini} P., 2011,
  \mnras, 414, L41

\bibitem[{{Galloway} {et~al}\mbox{.}(2008){Galloway}, {Muno}, {Hartman},
  {Psaltis}, \& {Chakrabarty}}]{2008ApJS..179..360G}
{Galloway} D.~K., {Muno} M.~P., {Hartman} J.~M., {Psaltis} D., {Chakrabarty}
  D., 2008, \apjs, 179, 360

\bibitem[{{Garc{\'{\i}}a} {et~al}\mbox{.}(2013){Garc{\'{\i}}a}, {Dauser},
  {Reynolds}, {Kallman}, {McClintock}, {Wilms}, \&
  {Eikmann}}]{2013ApJ...768..146G}
{Garc{\'{\i}}a} J., {Dauser} T., {Reynolds} C.~S., {Kallman} T.~R.,
  {McClintock} J.~E., {Wilms} J., {Eikmann} W., 2013, \apj, 768, 146

\bibitem[{{Garc{\'{\i}}a} \& {Kallman}(2010)}]{2010ApJ...718..695G}
{Garc{\'{\i}}a} J., {Kallman} T.~R., 2010, \apj, 718, 695

\bibitem[{{Gilfanov}(2010)}]{2010LNP...794...17G}
{Gilfanov} M., 2010, in Lecture Notes in Physics, Berlin Springer Verlag, Vol.
  794, Lecture Notes in Physics, Berlin Springer Verlag, {Belloni} T., ed.,
  p.~17

\bibitem[{{Gilfanov} {et~al}\mbox{.}(1998){Gilfanov}, {Revnivtsev}, {Sunyaev},
  \& {Churazov}}]{1998A&A...338L..83G}
{Gilfanov} M., {Revnivtsev} M., {Sunyaev} R., {Churazov} E., 1998, \aap, 338,
  L83

\bibitem[{{Gottwald} {et~al}\mbox{.}(1989){Gottwald}, {Haberl}, {Langmeier},
  {Hasinger}, {Lewin}, \& {van Paradijs}}]{1989ApJ...339.1044G}
{Gottwald} M., {Haberl} F., {Langmeier} A., {Hasinger} G., {Lewin} W.~H.~G.,
  {van Paradijs} J., 1989, \apj, 339, 1044

\bibitem[{{Grevesse} \& {Sauval}(1998)}]{1998SSRv...85..161G}
{Grevesse} N., {Sauval} A.~J., 1998, \ssr, 85, 161

\bibitem[{{Grindlay} {et~al}\mbox{.}(1976){Grindlay}, {Gursky}, {Schnopper},
  {Parsignault}, {Heise}, {Brinkman}, \& {Schrijver}}]{1976ApJ...205L.127G}
{Grindlay} J., {Gursky} H., {Schnopper} H., {Parsignault} D.~R., {Heise} J.,
  {Brinkman} A.~C., {Schrijver} J., 1976, \apjl, 205, L127

\bibitem[{{Hansen} \& {van Horn}(1975)}]{1975ApJ...195..735H}
{Hansen} C.~J., {van Horn} H.~M., 1975, \apj, 195, 735

\bibitem[{{Homer} {et~al}\mbox{.}(2001){Homer}, {Anderson}, {Margon},
  {Deutsch}, \& {Downes}}]{2001ApJ...550L.155H}
{Homer} L., {Anderson} S.~F., {Margon} B., {Deutsch} E.~W., {Downes} R.~A.,
  2001, \apjl, 550, L155

\bibitem[{{Iben}, {Tutukov} \& {Yungelson}(1995){Iben}, {Tutukov}, \&
  {Yungelson}}]{1995ApJS..100..233I}
{Iben}, Jr. I., {Tutukov} A.~V., {Yungelson} L.~R., 1995, \apjs, 100, 233

\bibitem[{{in't Zand} {et~al}\mbox{.}(2005){in't Zand}, {Cumming}, {van der
  Sluys}, {Verbunt}, \& {Pols}}]{2005A&A...441..675I}
{in't Zand} J.~J.~M., {Cumming} A., {van der Sluys} M.~V., {Verbunt} F., {Pols}
  O.~R., 2005, \aap, 441, 675

\bibitem[{{Jonker} \& {Nelemans}(2004)}]{2004MNRAS.354..355J}
{Jonker} P.~G., {Nelemans} G., 2004, \mnras, 354, 355

\bibitem[{{Jonker} {et~al}\mbox{.}(2001){Jonker}, {van der Klis}, {Homan},
  {M{\'e}ndez}, {van Paradijs}, {Belloni}, {Kouveliotou}, {Lewin}, \&
  {Ford}}]{2001ApJ...553..335J}
{Jonker} P.~G. {et~al.}, 2001, \apj, 553, 335

\bibitem[{{Juett} \& {Chakrabarty}(2003)}]{2003ApJ...599..498J}
{Juett} A.~M., {Chakrabarty} D., 2003, \apj, 599, 498

\bibitem[{{Juett} \& {Chakrabarty}(2005)}]{2005ApJ...627..926J}
{Juett} A.~M., {Chakrabarty} D., 2005, \apj, 627, 926

\bibitem[{{Juett} \& {Chakrabarty}(2006)}]{2006ApJ...646..493J}
{Juett} A.~M., {Chakrabarty} D., 2006, \apj, 646, 493

\bibitem[{{Juett}, {Psaltis} \& {Chakrabarty}(2001){Juett}, {Psaltis}, \&
  {Chakrabarty}}]{2001ApJ...560L..59J}
{Juett} A.~M., {Psaltis} D., {Chakrabarty} D., 2001, \apjl, 560, L59

\bibitem[{{Koliopanos}, {Gilfanov} \& {Bildsten}(2013){Koliopanos}, {Gilfanov},
  \& {Bildsten}}]{2013MNRAS.432.1264K}
{Koliopanos} F., {Gilfanov} M., {Bildsten} L., 2013, \mnras, 432, 1264

\bibitem[{{Kuulkers} {et~al}\mbox{.}(2003){Kuulkers}, {den Hartog}, {in't
  Zand}, {Verbunt}, {Harris}, \& {Cocchi}}]{2003A&A...399..663K}
{Kuulkers} E., {den Hartog} P.~R., {in't Zand} J.~J.~M., {Verbunt} F.~W.~M.,
  {Harris} W.~E., {Cocchi} M., 2003, \aap, 399, 663

\bibitem[{{Kuulkers} {et~al}\mbox{.}(2010){Kuulkers}, {in't Zand}, {Atteia},
  {Levine}, {Brandt}, {Smith}, {Linares}, {Falanga},
  {S{\'a}nchez-Fern{\'a}ndez}, {Markwardt}, {Strohmayer}, {Cumming}, \&
  {Suzuki}}]{2010A&A...514A..65K}
{Kuulkers} E. {et~al.}, 2010, \aap, 514, A65

\bibitem[{{Laor}(1991)}]{1991ApJ...376...90L}
{Laor} A., 1991, \apj, 376, 90

\bibitem[{{Linares} {et~al}\mbox{.}(2012){Linares}, {Connaughton}, {Jenke},
  {van der Horst}, {Camero-Arranz}, {Kouveliotou}, {Chakrabarty}, {Beklen},
  {Bhat}, {Briggs}, {Finger}, {Paciesas}, {Preece}, {von Kienlin}, \&
  {Wilson-Hodge}}]{2012ApJ...760..133L}
{Linares} M. {et~al.}, 2012, \apj, 760, 133

\bibitem[{{Madej} {et~al}\mbox{.}(2014){Madej}, {Garcia}, {Jonker}, {Parker},
  {Ross}, {Fabian}, \& {Chenevez}}]{2014arXiv1403.1432M}
{Madej} O.~K., {Garcia} J., {Jonker} P.~G., {Parker} M.~L., {Ross} R., {Fabian}
  A.~C., {Chenevez} J., 2014, ArXiv e-prints

\bibitem[{{Madej} {et~al}\mbox{.}(2010){Madej}, {Jonker}, {Fabian}, {Pinto},
  {Verbunt}, \& {de Plaa}}]{2010MNRAS.407L..11M}
{Madej} O.~K., {Jonker} P.~G., {Fabian} A.~C., {Pinto} C., {Verbunt} F., {de
  Plaa} J., 2010, \mnras, 407, L11

\bibitem[{{Markwardt}, {Juda} \& {Swank}(2003){Markwardt}, {Juda}, \&
  {Swank}}]{2003ATel..127....1M}
{Markwardt} C.~B., {Juda} M., {Swank} J.~H., 2003, The Astronomer's Telegram,
  127, 1

\bibitem[{{Nelemans} {et~al}\mbox{.}(2004){Nelemans}, {Jonker}, {Marsh}, \&
  {van der Klis}}]{2004MNRAS.348L...7N}
{Nelemans} G., {Jonker} P.~G., {Marsh} T.~R., {van der Klis} M., 2004, \mnras,
  348, L7

\bibitem[{{Nelemans}, {Jonker} \& {Steeghs}(2006){Nelemans}, {Jonker}, \&
  {Steeghs}}]{2006MNRAS.370..255N}
{Nelemans} G., {Jonker} P.~G., {Steeghs} D., 2006, \mnras, 370, 255

\bibitem[{{Nelson}, {Rappaport} \& {Joss}(1986){Nelson}, {Rappaport}, \&
  {Joss}}]{1986ApJ...311..226N}
{Nelson} L.~A., {Rappaport} S.~A., {Joss} P.~C., 1986, \apj, 311, 226

\bibitem[{{Ng} {et~al}\mbox{.}(2010){Ng}, {D{\'{\i}}az Trigo}, {Cadolle Bel},
  \& {Migliari}}]{2010A&A...522A..96N}
{Ng} C., {D{\'{\i}}az Trigo} M., {Cadolle Bel} M., {Migliari} S., 2010, \aap,
  522, A96

\bibitem[{{Paerels} {et~al}\mbox{.}(2001){Paerels}, {Brinkman}, {van der Meer},
  {Kaastra}, {Kuulkers}, {den Boggende}, {Predehl}, {Drake}, {Kahn}, {Savin},
  \& {McLaughlin}}]{2001ApJ...546..338P}
{Paerels} F. {et~al.}, 2001, \apj, 546, 338

\bibitem[{{Papitto} {et~al}\mbox{.}(2009){Papitto}, {Di Salvo}, {D'A{\`i}},
  {Iaria}, {Burderi}, {Riggio}, {Menna}, \& {Robba}}]{2009A&A...493L..39P}
{Papitto} A., {Di Salvo} T., {D'A{\`i}} A., {Iaria} R., {Burderi} L., {Riggio}
  A., {Menna} M.~T., {Robba} N.~R., 2009, \aap, 493, L39

\bibitem[{{Piraino} {et~al}\mbox{.}(1999){Piraino}, {Santangelo}, {Ford}, \&
  {Kaaret}}]{1999A&A...349L..77P}
{Piraino} S., {Santangelo} A., {Ford} E.~C., {Kaaret} P., 1999, \aap, 349, L77

\bibitem[{{Podsiadlowski}, {Rappaport} \& {Pfahl}(2002){Podsiadlowski},
  {Rappaport}, \& {Pfahl}}]{2002ApJ...565.1107P}
{Podsiadlowski} P., {Rappaport} S., {Pfahl} E.~D., 2002, \apj, 565, 1107

\bibitem[{{Rappaport} \& {Joss}(1984)}]{1984ApJ...283..232R}
{Rappaport} S., {Joss} P.~C., 1984, \apj, 283, 232

\bibitem[{{Savonije}, {de Kool} \& {van den Heuvel}(1986){Savonije}, {de Kool},
  \& {van den Heuvel}}]{1986A&A...155...51S}
{Savonije} G.~J., {de Kool} M., {van den Heuvel} E.~P.~J., 1986, \aap, 155, 51

\bibitem[{{Schulz} {et~al}\mbox{.}(2010){Schulz}, {Nowak}, {Chakrabarty}, \&
  {Canizares}}]{2010ApJ...725.2417S}
{Schulz} N.~S., {Nowak} M.~A., {Chakrabarty} D., {Canizares} C.~R., 2010, \apj,
  725, 2417

\bibitem[{{Shahbaz} {et~al}\mbox{.}(2008){Shahbaz}, {Watson}, {Zurita},
  {Villaver}, \& {Hernandez-Peralta}}]{2008PASP..120..848S}
{Shahbaz} T., {Watson} C.~A., {Zurita} C., {Villaver} E., {Hernandez-Peralta}
  H., 2008, \pasp, 120, 848

\bibitem[{{Smale} {et~al}\mbox{.}(1988){Smale}, {Mason}, {White}, \&
  {Gottwald}}]{1988MNRAS.232..647S}
{Smale} A.~P., {Mason} K.~O., {White} N.~E., {Gottwald} M., 1988, \mnras, 232,
  647

\bibitem[{{Strohmayer} \& {Bildsten}(2006)}]{2006csxs.book..113S}
{Strohmayer} T., {Bildsten} L., 2006, {New views of thermonuclear bursts},
  {Lewin} W.~H.~G., {van der Klis} M., eds., pp. 113--156

\bibitem[{{Tutukov} \& {Yungelson}(1993)}]{1993ARep...37..411T}
{Tutukov} A.~V., {Yungelson} L.~R., 1993, Astronomy Reports, 37, 411

\bibitem[{{van Haaften}, {Voss} \& {Nelemans}(2012){van Haaften}, {Voss}, \&
  {Nelemans}}]{2012A&A...543A.121V}
{van Haaften} L.~M., {Voss} R., {Nelemans} G., 2012, \aap, 543, A121

\bibitem[{{van Paradijs} \& {McClintock}(1994)}]{1994A&A...290..133V}
{van Paradijs} J., {McClintock} J.~E., 1994, \aap, 290, 133

\bibitem[{{Verbunt} \& {van den Heuvel}(1995)}]{1995xrbi.nasa..457V}
{Verbunt} F., {van den Heuvel} E.~P.~J., 1995, X-ray Binaries, 457

\bibitem[{{Walter} {et~al}\mbox{.}(1982){Walter}, {Mason}, {Clarke}, {Halpern},
  {Grindlay}, {Bowyer}, \& {Henry}}]{1982ApJ...253L..67W}
{Walter} F.~M., {Mason} K.~O., {Clarke} J.~T., {Halpern} J., {Grindlay} J.~E.,
  {Bowyer} S., {Henry} J.~P., 1982, \apjl, 253, L67

\bibitem[{{Werner} {et~al}\mbox{.}(2006){Werner}, {Nagel}, {Rauch}, {Hammer},
  \& {Dreizler}}]{2006A&A...450..725W}
{Werner} K., {Nagel} T., {Rauch} T., {Hammer} N.~J., {Dreizler} S., 2006, \aap,
  450, 725

\bibitem[{{Wijnands} \& {van der Klis}(1998)}]{1998Natur.394..344W}
{Wijnands} R., {van der Klis} M., 1998, \nat, 394, 344

\bibitem[{{Yungelson}, {Nelemans} \& {van den Heuvel}(2002){Yungelson},
  {Nelemans}, \& {van den Heuvel}}]{2002A&A...388..546Y}
{Yungelson} L.~R., {Nelemans} G., {van den Heuvel} E.~P.~J., 2002, \aap, 388,
  546

\bibitem[{{Zhong} \& {Wang}(2011)}]{2011ApJ...729....8Z}
{Zhong} J., {Wang} Z., 2011, \apj, 729, 8

\end{thebibliography}

\label{lastpage}
\end{document}